\documentclass[preprint]{aastex63}
\usepackage{mathtools}
\usepackage{color}
\usepackage{multirow}
\usepackage{comment}

\turnoffeditone

\usepackage{bm}
\usepackage{here}
\usepackage{textcomp} 
\usepackage{physics}
\usepackage{comment}
\usepackage{mathtools}

\newcommand{\mr}{\mathrm}
\renewcommand{\thefootnote}{*\arabic{footnote}}

\newcommand{\Equref}[1]{Equation~(\ref{#1})}
\newcommand{\Figref}[1]{Figure~\ref{#1}}
\newcommand{\Secref}[1]{\S~\ref{#1}}

\newcommand{\Equrefs}[1]{Equations~(\ref{#1})}
\newcommand{\Figrefs}[1]{Figures~\ref{#1}}

\received{February 15, 2022}
\revised{June 21, 2022}
\accepted{June 30, 2022}
\submitjournal{Astrophysical Journal}

\shorttitle{Accretion of Gas onto Circumplanetary Disk}
\shortauthors{Maeda et al.}
\graphicspath{{./}{figures/}}

\begin{document}

\title{Delivery of gas onto the circumplanetary disk of giant planets: Planetary-mass dependence of the source region of accreting gas and mass accretion rate}

\correspondingauthor{Natsuho Maeda}
\email{nmaeda@stu.kobe-u.ac.jp}

\author[0000-0002-6874-5178]{Natsuho Maeda}
\affiliation{Department of Planetology, Kobe University, 1-1 Rokkodai-cho, Nada-ku, Kobe 657-8501, Japan}

\author[0000-0002-4383-8247]{Keiji Ohtsuki}
\affiliation{Department of Planetology, Kobe University, 1-1 Rokkodai-cho, Nada-ku, Kobe 657-8501, Japan}

\author[0000-0002-5964-1975]{Takayuki Tanigawa}
\affiliation{National Institute of Technology, Ichinoseki College,
Ichinoseki 021-8511, Japan}

\author[0000-0002-0963-0872]{Masahiro N. Machida}
\affiliation{Department of Earth and Planetary Science, Graduate School of Science, Kyushu University, 
Fukuoka 819-0395, Japan}

\author[0000-0002-8478-1881]{Ryo Suetsugu}
\affiliation{National Institute of Technology, Oshima College, Oshima 742-2193, Japan}







\begin{abstract}
Gas accretion onto the circumplanetary disks and the source region of accreting gas are important to reveal dust accretion that leads to satellite formation around giant planets.
We performed local three-dimensional high-resolution hydrodynamic simulations of isothermal and inviscid gas flow around a planet to investigate planetary-mass dependence of gas accretion band width and gas accretion rate onto circumplanetary disks.
We examined cases with various planetary masses corresponding to $M_{\mr{p}}=0.05-1M_{\mr{Jup}}$ at 5.2~au, where $M_{\mr{Jup}}$ is the current Jovian mass.
We found that the radial width of the gas accretion band is proportional to $M_{\mr{p}}^{1/6}$ for the low-mass regime with $M_{\mr{p}} \lesssim 0.2 M_{\mr{Jup}}$ while it is proportional to $M_{\mr{p}}$ for the high-mass regime with $M_{\mr{p}}\gtrsim 0.2M_{\mr{Jup}}$. 
We found that the ratio of the mass accretion rate onto the circumplanetary disk to that into the Hill sphere is about 0.4 regardless of planetary mass for the cases we examined.
Combining our results with the gap model obtained from global hydrodynamic simulations, we derive semi-analytical formulae of mass accretion rate onto circumplanetary disks. We found that the mass dependence of our three-dimensional accretion rates is the same as the previously-obtained two-dimensional case, although the qualitative behavior of accretion flow onto the CPD is quite different between the two cases.
\end{abstract}



\section{INTRODUCTION} \label{sec:intro}

Recent developments of numerical hydrodynamic simulations have made progress in understanding gas flow around protoplanets. 
When a protoplanet accretes gas from the protoplanetary disk (PPD), a circumplanetary disk (CPD) naturally forms around a sufficiently massive protoplanet while a spherically symmetric envelope forms around a small protoplanet  \citep[e.g.,][]{tw02,m08,m09,ab09,t12,g13,sz14,sz16,sz17,f19,l19,s19,s20}.
Whether envelopes or CPDs form depends not only on the planetary mass but also on the surface temperature of the planet and the equation of state \citep{sz16,sz17,f19}.
High-resolution three-dimensional hydrodynamic simulations demonstrate a meridional circulation of the gas near a gap opened by a giant planet in the PPD: gas flows into the gap at a high altitude over the CPD, and the planet pushes the gas back into the PPD while accreting part of the gas \citep[][]{t12,sz14,m14,fc16}.
Similar circulating flows have also been reported in simulations around less massive planets corresponding to super-Earths and sub-Neptunes \citep{o15,f15,c17,kt18,k19}. 

Growing giant planets accumulate gas and solid particles via their CPD, and principal regular satellites, such as the Galilean satellites and Titan, are thought to be formed in the CPD.
Thus, accretion of gas and solid particles onto the CPD is important for both planet and satellite formation.
Kilometer-sized or larger planetesimals can be captured by gas drag from the CPD \citep{f13}. Some of the captured planetesimals can be ablated and become particles with the size of pebbles or dust, and the remaining part can become satellitesimals \citep[][]{rj20}.
However, capture efficiency of planetesimals significantly decreases when a gap in the planetesimal disk forms at the planet's orbit and/or the velocity dispersion of the planetesimals is too large \citep[][]{f13,s16,so17}, unless the planetesimal reservoir exterior to the gap is perturbed by other planetary embryos \citep[][]{r18}.
On the other hand, dust particles accreted with gas are also potential building blocks of satellites and would become a major contributor in circumstances where planetesimal capture is inefficient \citep[][]{cw02,cw06,so17,sh19,bm20}.

The motion of sufficiently small particles delivered from the PPD into the CPD is strongly influenced by the gas flow around the planet, such as the meridional circulation mentioned above.
For example, using high-resolution local three-dimensional hydrodynamic simulations assuming isothermal and inviscid gas around a growing giant planet, \citet{t12} found that the gas vertically accretes onto the CPD but flows radially outward in the midplane, while the direction of the radial flow in the midplane of the CPD would depend on the viscosity of the gas \citep[][]{sz14}.
Using orbital integration of particles that takes account of the gas flow obtained by \citet{t12}, \citet{t14} examined accretion of particles initially confined within the midplane of the PPD. They found that mm-sized or smaller particles (i.e., those with the Stokes number significantly smaller than unity) cannot accrete into the CPD because of the strong influence of the outflowing gas in the midplane. On the other hand, \citet{h20} performed similar calculations but took account of the vertical motions of the gas and particles. They found that sufficiently small particles coupled to the vertically accreting gas can accrete onto the CPD when they are vertically stirred up in the PPD.
However, they also found that the dust-to-gas ratio in the gas accreting onto the CPD can become significantly smaller than that in the PPD if the turbulence in the PPD is weak 
\citep[corresponding to the turbulence parameter $\alpha \sim 10^{-4}$,][]{ss73}.
More recently, \citet{bi21} and \citet{sz22} carried out global three-dimensional hydrodynamic simulations with dust particles. They found that the dust particles are stirred up vertically well above the midplane due to the meridional circulation, where the spiral wakes created by the planet bring up the dust from the midplane.

Thus, it is crucial for understanding satellite formation around gas giants to study the delivery of gas and dust particles onto the CPD from the PPD using three-dimensional high-resolution hydrodynamic simulations, such as the one shown in \citet{t12}.
However, \citet{t12} (and also Tanigawa et al. 2014 and Homma et al. 2020) examined only the case of $r_{\mr{H}}/h_{\mr{g}}=1$, where $r_{\mr{H}}$ is the planet's Hill radius and $h_{\mr{g}}$ is the scale height of the gas in the PPD, corresponding to a planetary-mass of $0.4M_{\mr{Jup}}$ at the current Jovian orbit for the minimum-mass solar nebula model \citep[MMSN;][]{h81}.
Since formation of principal regular satellites is likely to occur during planet formation, it is necessary to understand how the flow pattern and amount of gas and dust particles delivered onto the CPD depend on planetary mass. 
The dependence of gas and dust supply on planetary mass will also help us understand satellite formation in different planet systems, such as Jupiter and Saturn, or exoplanet systems.

In a series of papers, we will investigate planetary-mass dependence of delivery of gas and dust particles onto CPDs. 
As the first step of understanding planetary-mass dependence of dust supply onto CPDs, we focus on gas accreting onto CPDs in this paper.
Accretion of dust particles that are not completely coupled to the gas will be investigated in our subsequent paper.
Details of the structure of the CPD depend not only on the planetary mass but also on various other factors such as the equation of state, viscosity, and surface temperature of the planet \citep{m09,m10,sz14,sz16,sz17,g13,f19}. 
In contrast, the picture that the gas around a sufficiently massive planet vertically accretes onto the CPD is rather insensitive to numerical settings \citep{m10,sz14,sz17,g13}. 
Thus, we consider the simplest case of an isothermal and inviscid fluid, following \citet{t12}. 
We perform three-dimensional high-resolution hydrodynamic simulations with various planetary masses and investigate the dependence of gas accretion on planetary mass. 

The structure of this paper is as follows. In \Secref{sec:methods}, we will explain basic equations and numerical settings. In \Secref{sec:results}, we will show the results of hydrodynamic simulations. In \Secref{sec:discussions}, we will derive the semi-analytical formulae of mass accretion rate of gas by combining our results with the gap model obtained from global hydrodynamic simulations \citep{k15}. We will also discuss comparisons with previous works in \Secref{sec:discussions}. We will summarize this work in \Secref{sec:conclusions}.

\section{Methods}\label{sec:methods}
\subsection{Basic equations}
The model used in this work is similar to those described in \cite{m08} and \cite{t12}.
We consider a giant planet on a circular orbit embedded in a PPD.
We erect a rotating local Cartesian coordinate system with origin at the planet, the $x$-axis pointing radially outward, the $y$-axis in the direction of the planet's orbital motion, and the $z$-axis in the vertical direction to the PPD midplane. We define $r$ and $R$ as $r\equiv \sqrt{x^2+y^2+z^2}$ and $R\equiv \sqrt{x^2+y^2}$, respectively.

As in \citet{t12}, we assume that the gas is inviscid and isothermal, and denote its density, velocity, and pressure as $\rho$, $\bm{v}$, and $P$, respectively. We normalize time by the inverse of the Keplerian angular velocity $\Omega_{\mr{K}}^{-1}=(GM_{\mr{c}}/a^3)^{-1/2}$, length by the scale height of the PPD $h_{\mr{g}}$, and density by $\Sigma_0/h_{\mr{g}}$, where $G$, $M_{\mr{c}}$, $a$, and $\Sigma_0$ are the gravitational constant, mass of the central star, semi-major axis of the planet, and unperturbed surface density of the PPD, respectively. In this case, velocity is normalized by sound speed $c_{\mr{s}}=h_{\mr{g}}\Omega_{\mr{K}}$. In the following, hats denote normalized quantities. Then, the basic equations can be written as \citep[for details, see][]{m08}
\begin{eqnarray}\label{eq:basicEq}
\frac{\partial \hat{\rho}}{\partial \hat{t}} + \hat{\nabla} \cdot (\hat{\rho} \hat{\bm{v}})  &=& 0,\\
\frac{\partial \hat{\bm{v}}}{\partial \hat{t}} + (\hat{\bm{v}} \cdot \hat{\nabla})\hat{\bm{v}}  &=& -\frac{1}{\hat{\rho}}\hat{\nabla} \hat{P}  -\hat{\nabla} \hat{\Phi} - 2 \bm{e}_{z}\times \hat{\bm{v}},\\
\hat{P} &=& \hat{\rho},
\end{eqnarray}
where $\bm{e}_{z}=(0,0,1)$ is the unit vector in the $z$-direction, and 
\begin{eqnarray}
\hat{\Phi}&=&\hat{\Phi}_{\mr{tidal}} + \hat{\Phi}_{\mr{p}} + \frac{9}{2}\hat{r}_{\mr{H}}^2, \label{eq:Phi}\\
\hat{\Phi}_{\mr{tidal}} &=& -\frac{3}{2}\hat{x}^2 + \frac{1}{2}\hat{z}^2,  \label{eq:Phi_t}\\
\hat{\Phi}_{\mr{p}} &=& -\frac{3\hat{r}_{\mr{H}}^3}{\hat{r}}. \label{eq:Phi_p}
\end{eqnarray}
In the above, $\hat{\Phi}_{\mr{tidal}}$ is the tidal potential of the central star, $\hat{\Phi}_{\mr{p}}$ is the gravitational potential of the planet, and a constant $(9/2)\hat{r}_{\mr{H}}^2$ is added in \Equref{eq:Phi} so that $\hat{\Phi}=0$ at the two Lagrangian points $\mr{L_1}$ and $\mr{L_2}$ of the planet;
$\hat{r}_{\mr{H}}$ is the normalized Hill radius of the planet denoted by
\begin{equation}\label{eq:hat_rH}
    \hat{r}_{\mr{H}}=\frac{r_{\mr{H}}}{h_{\mr{g}}} = \left( \frac{M_{\mr{p}}}{3M_{\mr{c}}} \right)^{1/3} \frac{a}{h_{\mr{g}}},
\end{equation}
where $M_{\mr{p}}$ is the mass of the planet. This parameter $\hat{r}_{\mathrm{H}}$ is the only one physical parameter in our formalization.
Note that $\hat{r}_{\rm H}=1$ corresponds to the planetary mass $M_{\rm p}$ equal to the so-called thermal mass, $3(h_{\rm g}/a)^3 M_{\rm c}$ \citep[e.g.,][]{gc19}

To determine the quantitative relationship between $\hat{r}_{\mr{H}}$ and $M_{\mr{p}}$, we need to give a semimajor axis dependence of the disk scale height. If we consider an optically thin disk and assume that the temperature distribution is $T=280\; \mr{K} \; (a/1\mr{\;au})^{-1/2}$ \citep{h81}, the scale height can be derived as
\begin{equation}\label{eq:hayashi_disk}
h_{\mr{g}}= c_{\mr{s}} \Omega_{\mr{K}}^{-1} = 4.93 \times 10^9 \left( \frac{a}{1\mr{\; au}} \right)^{5/4} \;\mr{m}.
\end{equation}
Using \Equrefs{eq:hat_rH} and (\ref{eq:hayashi_disk}), we obtain
\begin{equation}\label{eq:mp_hayashi}
\frac{M_{\mr{p}}}{M_{\mr{Jup}}} = 0.41 \left( \frac{M_{\mr{c}}}{1M_{\odot}} \right)^{1/2} \left( \frac{a}{5.2 \mr{\;au}} \right)^{3/4} \hat{r}_{\mr{H}}^3.
\end{equation}
On the other hand, recent MHD simulations show that the temperature in magnetically accreting disks is lower than that in viscous accretion disks \citep[][]{m19}.
If we consider a colder disk and assume that the temperature distribution is $T=100\; \mr{K} \; (a/1\mr{\;au})^{-3/7}$ \citep{k70}, the scale height can be derived as 
\begin{equation}\label{eq:kusaka_disk}
h_{\mr{g}}=c_{\mr{s}} \Omega_{\mr{K}}^{-1} = 2.93 \times 10^9 \left( \frac{a}{1\mr{\; au}} \right)^{9/7} \;\mr{m},
\end{equation}
and from \Equrefs{eq:hat_rH} and (\ref{eq:kusaka_disk}), we obtain
\begin{equation}\label{eq:mp_kusaka}
\frac{M_{\mr{p}}}{M_{\mr{Jup}}} = 0.098 \left( \frac{M_{\mr{c}}}{1M_{\odot}} \right)^{3/7} \left( \frac{a}{5.2 \mr{\;au}} \right)^{6/7} \hat{r}_{\mr{H}}^3.
\end{equation}

\begin{figure}[H]
	\begin{center}
		\includegraphics*[bb=0 0 343 241,scale=0.9]{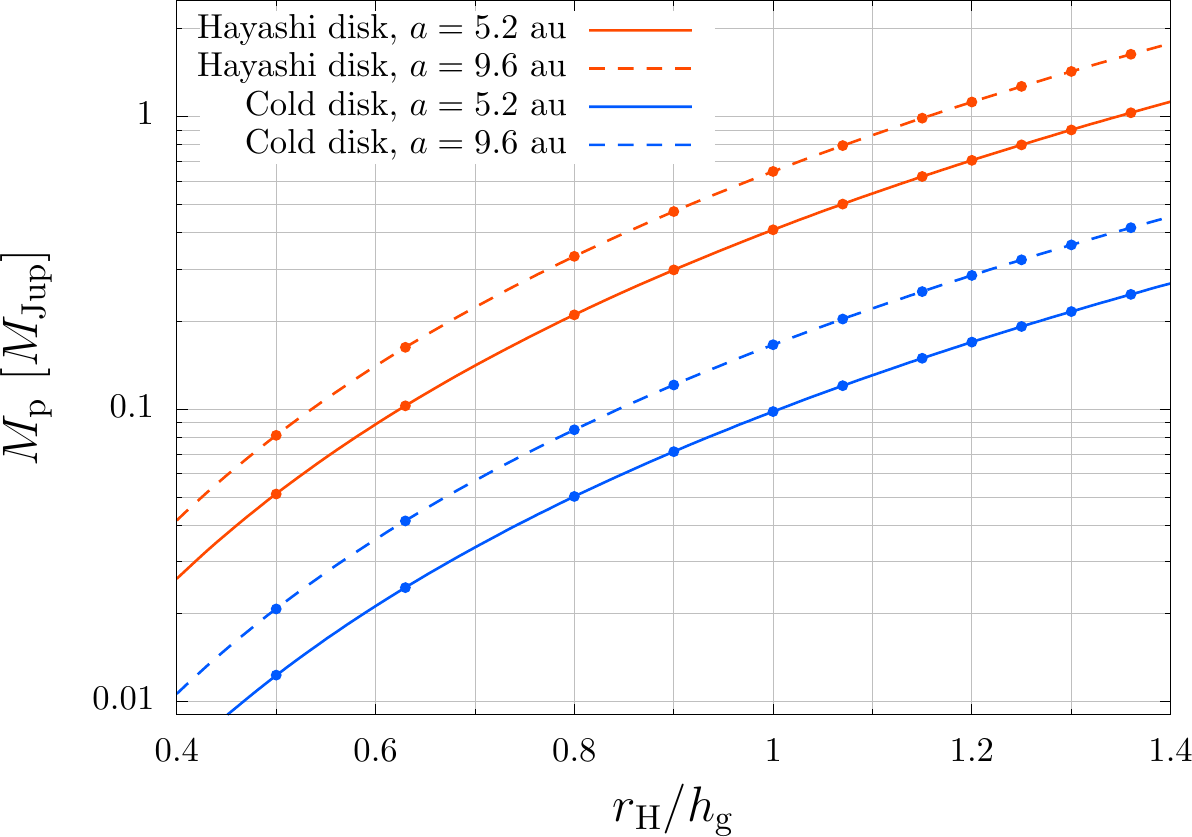}
	\end{center}
	\caption{Relationship between $\hat{r}_{\mr{H}}=r_{\mr{H}}/h_{\mr{g}}$ and $M_{\mr{p}}$. The red lines are drawn from \Equref{eq:mp_hayashi}, which corresponds to the Hayashi model ($T=280\; \mr{K} \;(a/1\mr{\;au})^{-1/2}$). The blue lines are drawn from \Equref{eq:mp_kusaka}, which corresponds to the cold disk (Kusaka) model ($T=100\; \mr{K} \;(a/1\mr{\;au})^{-3/7}$). The solid and dashed lines show the cases for $a=5.2$~au (the Jovian orbit) and 9.6~au (the Saturnian orbit), respectively. The symbols correspond to the values of $\hat{r}_{\rm H}$ we adopted in our simulations.}
	\label{fig:ratio-Mp}
\end{figure}

\Figref{fig:ratio-Mp} shows the relationship between $\hat{r}_{\mr{H}}$ and $M_{\mr{p}}$ (i.e., Equations (\ref{eq:mp_hayashi}) and (\ref{eq:mp_kusaka})) in the case of $a=5.2$~au and $a=9.6$~au, respectively, for the two disk models described above.
To investigate the dependence of gas accretion on planetary mass, we ran 11 simulations varying the value of $\hat{r}_{\mr{H}}$ and analyzed quasi-steady state for each run. 
We ran the simulations with 11 values of $\hat{r}_{\rm H}$ in the range of $\hat{r}_{\mr{H}}=0.5-1.36$, corresponding to $M_{\mr{p}}=0.05M_{\mr{Jup}}-1M_{\mr{Jup}}$ for the Hayashi disk (\Equref{eq:mp_hayashi}) and $M_{\mr{p}}=0.01M_{\mr{Jup}}-0.2M_{\mr{Jup}}$ for the cold disk (\Equref{eq:mp_kusaka}) at the current Jovian orbit, respectively.

In the following, when we refer to $M_{\mr{p}}$ assumed in each simulation, it is obtained from \Equref{eq:mp_hayashi} assuming $a=5.2$~au (the current Jovian orbit) and $M_{\mr{c}}=M_{\odot}$ (the solar mass), unless explicitly stated otherwise.

\newpage
\subsection{Numerical settings}
We used a three-dimensional nested-grid hydrodynamic simulation code \citep[e.g.,][]{m05,m06}. Our numerical settings are basically the same as in \cite{t12}. Assuming symmetry about the $\hat{z}=0$ plane, we set the sizes of the computational domain to $\hat{x}\in [-12, 12]$, $\hat{y}\in [-12, 12]$, and $\hat{z}\in [0, 6]$, i.e., the length of each side of the simulation box is $(\hat{L}_x,\; \hat{L}_y,\; \hat{L}_z)=(24,\; 24,\; 6)$.
We set the nested-grid composed of 11 levels at maximum ($l_{\mr{max}}=11$), where $l$ is the level of the nested-grid. As the level increases, the length of each side is halved keeping the planet at the origin, while the numbers of cells in each grid are kept constant as $(n_x,\; n_y,\; n_z)=(64,\; 64,\; 16)$. The finest cell width is $\hat{L}_x/64/2^{11-1}=3.66 \times 10^{-4}$, which roughly corresponds to one-fourth of the physical size of a planet with a mean density of 1~$\mr{g\,cm^{-3}}$.

Initial and boundary conditions are also the same as in \cite{t12}. Initially, the gas density distribution is in hydrostatic equilibrium in the $z$-direction, and the Keplerian velocity field is assumed. We set the unperturbed boundary in the $x$-direction, the mixed condition of unperturbed and periodic boundaries in the $y$-direction, and the mirror condition in the $z$-direction, respectively \citep[for details, see][]{t12}.
We set the smoothing length $r_{\mr{sm}}=2.44\times 10^{-4}h_{\mr{g}}$ to avoid singularity of the planet gravity, but no sink cells are set in the present work.

We set $l=1$ at the start of each simulation and add deeper levels of grids ($l>1$) gradually with time. We empirically determined to create a new level with $l=2$ at $\hat{t}=20$, $l=3$ at $\hat{t}=50$, and $l=4$ at $\hat{t}=150$, respectively, and deeper levels with $l\geq 5$ are created when the gas flow in each level is sufficiently relaxed \citep[see also][]{t12}.
We observed that non-negligible oscillations appeared in the density and velocity fields for the case of $\hat{r}_{\mr{H}}>1.0$ (i.e., $M_{\mr{p}}>0.4M_{\mr{Jup}}$), while quasi-stationary flow patterns can be obtained for $\hat{r}_{\mr{H}}\leq 1.0$. 
In order to analyze the field data, while we use snapshots at the end of each simulation for the case of $\hat{r}_{\mr{H}}\leq 1.0$, we obtain time-averaged values of the density and velocity for $\hat{r}_{\mr{H}}>1.0$ as described below.
For $l\leq 4$, we take a time average for $\hat{t}\geq 150$ because numerical oscillations mainly occur in the region of $l\leq 4$. We can reduce the effect of numerical oscillation by taking a time average of the snapshots over a sufficiently long duration compared to the period of the oscillation. For $l\geq 5$, we use snapshots at the end of each simulation ($\hat{t}\sim 170$). The spatial region corresponding to $l\geq5$ is roughly within the Hill sphere and the gas distribution is nearly axisymmetric, thus discontinuity in the gas distribution and flow patterns between this region and the other region with $l\leq4$ is negligible. Thus, we combine time-averaging data for $l\leq 4$ and snapshots for $l\geq 5$. 
We confirmed that the discontinuities of combined gas fields do not significantly affect our qualitative results, while this method might affect quantitative results to some degree, such as gas accretion rates \citep[][]{f19}.

\section{Results}\label{sec:results}
\subsection{Overview of gas fields}
\begin{figure}[h]
\begin{tabular}{cc}
	\begin{minipage}{0.5\hsize}	
		\begin{center}
			\includegraphics*[bb=0 0 901 1093,scale=0.25]{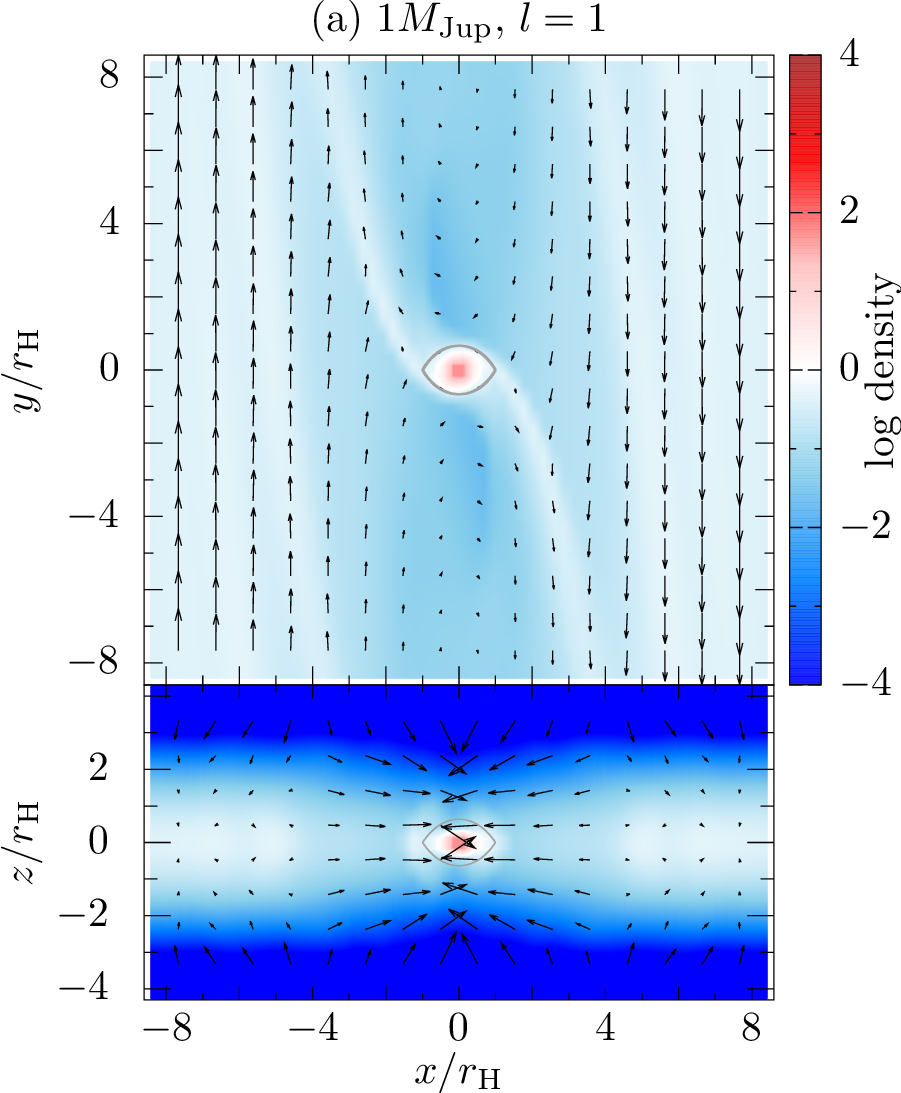}
		\end{center}
	\end{minipage}&
	
	\begin{minipage}{0.5\hsize}	
		\begin{center}
			\includegraphics*[bb=0 0 922 1103,scale=0.25]{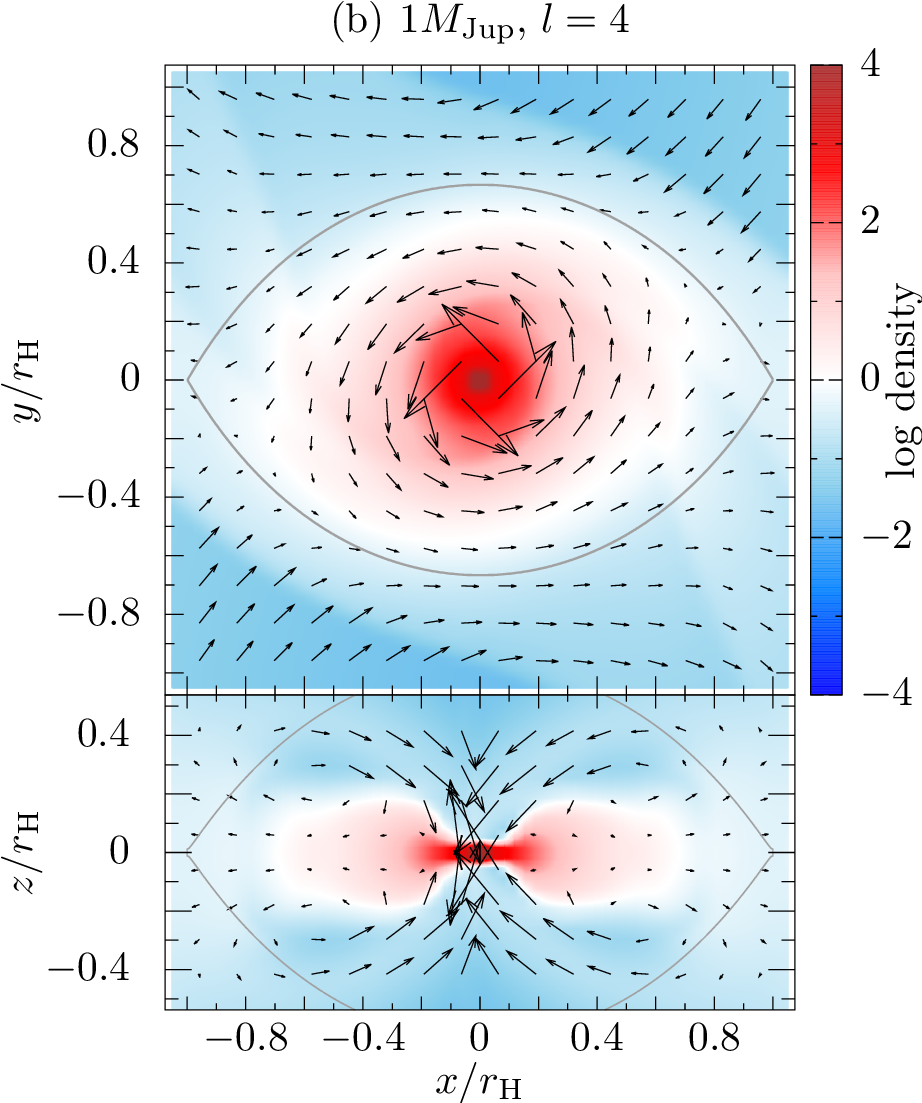}
		\end{center}
	\end{minipage}
\end{tabular}
\caption{Density distributions (color) and velocity vectors (arrows) around the planet ((a) for $l=1$, (b) for $l=4$) for $\hat{r}_{\mr{H}}=1.36$ ($M_{\mr{p}}=1M_{\mr{Jup}}$). The upper and lower panels correspond to the planes where the cell closest to $z=0$ and $y=0$, respectively. The lemon-shaped region enclosed by the gray line in each panel is the planet's Hill sphere (its surface is defined by $\hat{\Phi}=0$ in \Equref{eq:Phi}). Sizes of velocity vectors are appropriately chosen in each panel.}
	\label{fig:gasfield}
\end{figure}

\Figrefs{fig:gasfield}a and \ref{fig:gasfield}b show density distributions and velocity vectors for levels $l=1$ and 4, respectively, for the model with $\hat{r}_{\mr{H}}=1.36$ ($M_{\mr{p}}=1M_{\mr{Jup}}$) at $\hat{t}=170$. The upper and lower panels show the distributions for the planes where the cell closest to $z=0$ and $y=0$, respectively. In panel (a), two arm-like structures are extended from the planet's Hill sphere, and a gap region with low gas surface density is formed around the planet's orbit. The enlarged maps around the Hill spheres are shown in panel (b), where a CPD with high density and an axisymmetric structure can be confirmed within the Hill sphere. The vertical inflow from high altitude to the CPD along the $z$-axis and a weak outflow at the midplane can be observed in the lower panels.

\subsection{Structure of gas flow within the Hill sphere}
\begin{figure}[H]
	\begin{center}
		\includegraphics*[bb=0 0 251 245,scale=0.9]{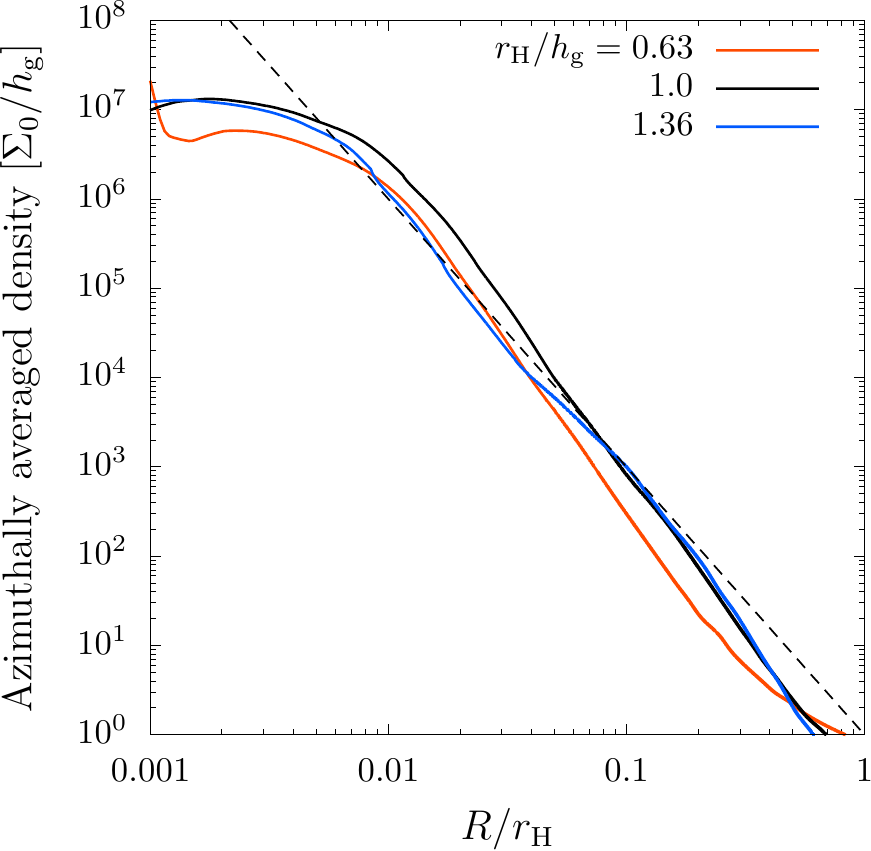}
	\end{center}
	\caption{Radial density distributions at the midplane as a function of the horizontal distance from the planet $R=\sqrt{x^2+y^2}$ scaled by the Hill radius $r_{\mr{H}}$. Density is azimuthally averaged around the planet. The red, black, and blue lines correspond to the cases of $\hat{r}_{\mr{H}}=0.63$ ($M_{\mr{p}}=0.1M_{\mr{Jup}}$), 1.0 ($M_{\mr{p}}=0.4M_{\mr{Jup}}$), and 1.36 ($M_{\mr{p}}=1M_{\mr{Jup}}$), respectively. The dashed line shows the slope for the power-law distribution of $\hat{\rho} \propto \hat{R}^{-3}$.}
	\label{fig:rho-fit}
\end{figure}

\Figref{fig:rho-fit} shows the radial distribution of the azimuthally averaged gas density at the midplane for the cases of $\hat{r}_{\mr{H}}=0.63$ ($M_{\mr{p}}=0.1M_{\mr{Jup}}$), 1.0 ($M_{\mr{p}}=0.4M_{\mr{Jup}}$), and 1.36 ($M_{\mr{p}}=1M_{\mr{Jup}}$).
We found that the density distributions within the Hill sphere approximately follow the power-law distribution of $\hat{\rho} \propto \hat{R}^{-3}$ for all cases, which is consistent with \cite{t12}.

\begin{figure}[H]
	\begin{center}
		\includegraphics*[bb=0 0 206 258,scale=0.9]{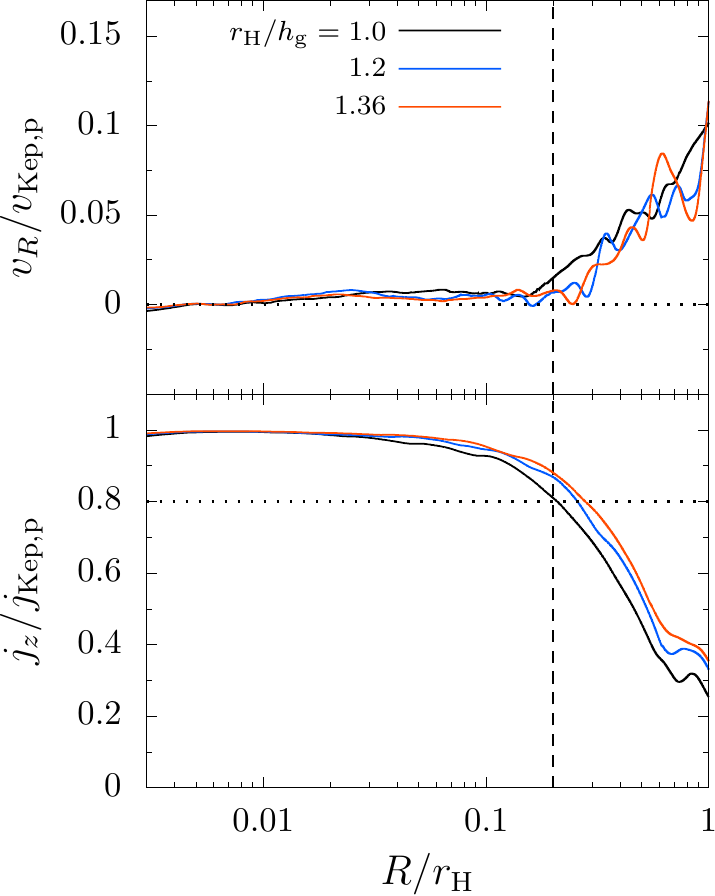}
	\end{center}
	\caption{Azimuthally averaged radial velocity (top) and specific angular momentum (bottom) around the planet at the midplane, normalized by the value for the Keplerian rotation around the planet. The black, blue, and red lines correspond to the cases of $\hat{r}_{\mr{H}}=1.0$ ($M_{\mr{p}}=0.4M_{\mr{Jup}}$), 1.2 ($M_{\mr{p}}=0.7M_{\mr{Jup}}$), and 1.36 ($M_{\mr{p}}=1M_{\mr{Jup}}$), respectively. The vertical dashed line represents the radius of the CPD, $R=0.2r_{\mr{H}}$. Rotational velocity of the gas at $R<0.2r_{\mr{H}}$ are faster than 80\% of the Keplerian velocity for all cases.}
	\label{fig:vr_high}
\end{figure}

When the planetary mass is high enough, radial structure of the gas flow within the Hill radius is characterized by the Hill radius.
\Figref{fig:vr_high} shows radial distributions of radial velocity $v_R$ (top) and the $z$-component of specific angular momentum $j_z$ (bottom) around the planet at the midplane for the high mass cases ($\hat{r}_{\mr{H}}=1.0 - 1.36$; $M_{\mr{p}}=0.4M_{\mr{Jup}} - 1M_{\mr{Jup}}$).\footnote{In Figure 7 of \citet{t12}, $v_R$ has positive values for $R<0.006h_{\rm g}$, while our result for $\hat{r}_{\rm H}=1.0$ has very small negative values for the same region. This discrepancy is caused by the difference of extrapolation methods. Extrapolation is needed to obtain the values at the midplane because physical quantities are calculated only for the $z>0$ regions in the hydrodynamic simulation. \citet{t12} used quadratic functions for the extrapolation in $z$-direction. Although the raw values of $v_R$ at the cells closest to the midplane are small negative values, non-negligible positive values can be artificially obtained when the points used for the extrapolation cross a discontinuity where higher-order extrapolations are not suitable, such as a surface of CPD.}
Both radial velocity and specific angular momentum are azimuthally averaged around the planet and normalized by the values for the Keplerian rotation around the planet ($v_{\mr{Kep,p}}$ and $j_{\mr{Kep,p}}$, respectively). Note that the specific angular momentum is measured in the rotational frame. 
We found that the radial distributions of the scaled radial velocity and the scaled specific angular momentum do not depend on planetary mass appreciably. 
We also found that gas elements at $R<0.2r_{\mr{H}}$ (left side of the vertical dashed line) rotate with a velocity over 80\% of the Keplerian velocity around the planet.
Defining this region as a CPD, the radius of the CPD is $0.2r_{\mr{H}}$, i.e., proportional to the Hill radius of the planet for $\hat{r}_{\mr{H}}\geq 1.0$ ($M_{\mr{p}}\geq 0.4M_{\mr{Jup}}$).
The CPD radius of our result is roughly consistent with previous works which examined cases for $M_{\rm p}\sim M_{\rm Jup}$ \citep{d03,ab09,ab12,t12,sz14}.

\begin{figure}[H]
	\begin{center}
		\includegraphics*[bb=0 0 206 258,scale=0.9]{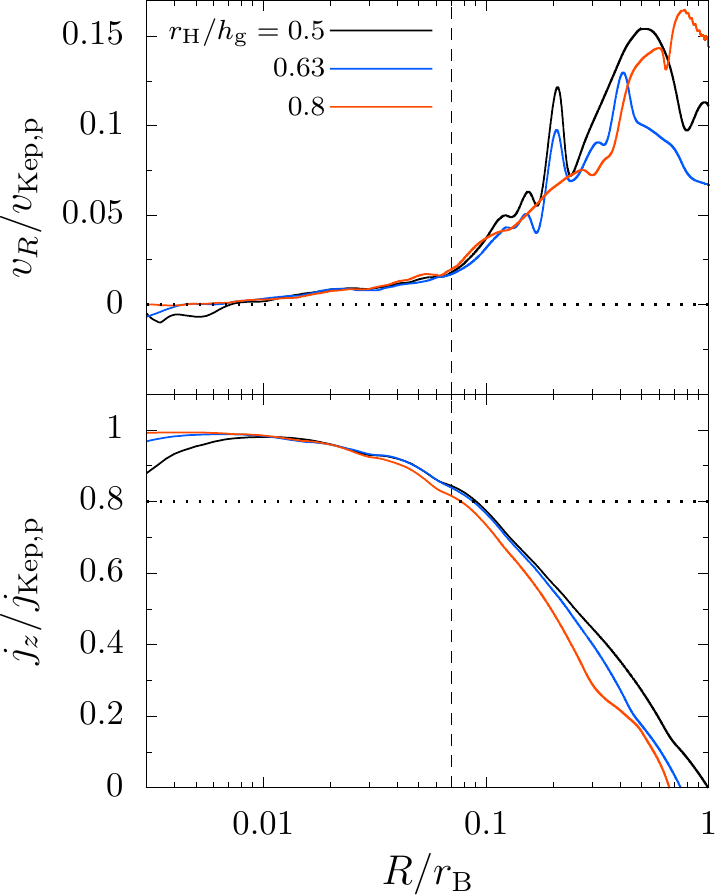}
	\end{center}
	\caption{Same as \Figref{fig:vr_high} but for the low planetary-mass case. The horizontal axis is normalized by the Bondi radius $r_{\mr{B}}$. The black, blue, and red lines correspond to the cases of $\hat{r}_{\mr{H}}=0.5$ ($M_{\mr{p}}=0.05M_{\mr{Jup}}$), 0.63 ($M_{\mr{p}}=0.1M_{\mr{Jup}}$), and 0.8 ($M_{\mr{p}}=0.2M_{\mr{Jup}}$), respectively. The vertical dashed line represents the radius of the CPD, $R=0.07r_{\mr{B}}$. Rotational velocity of the gas at $R<0.07r_{\mr{B}}$ is faster than 80\% of the Keplerian velocity for all cases.}
	\label{fig:vr_low}
\end{figure}

For the low mass cases, on the other hand, radial structure of the gas flow within the Hill radius is characterized by the Bondi radius $r_{\mr{B}}=GM_{\mr{p}}/c^2_{\mr{s}}$.
\Figref{fig:vr_low} shows radial distributions of $v_R$ and $j_z$ for the low mass cases ($\hat{r}_{\mr{H}}=0.5 - 0.8$; $M_{\mr{p}}=0.05M_{\mr{Jup}} - 0.2M_{\mr{Jup}}$). 
We found that $v_R$ and $j_z$ shown as a function of the horizontal radial distance scaled by the Bondi radius do not depend on planetary mass appreciably.
Gas elements at $R<0.07r_{\mr{B}}$ (left side of the vertical dashed line) rotate with a velocity over 80\% of the Keplerian velocity around the planet. Thus, the radius of the CPD in this case is proportional to the Bondi radius for $\hat{r}_{\mr{H}}\leq 0.8$ ($M_{\mr{p}}\leq 0.2M_{\mr{Jup}}$).

Our results that the radii of CPDs are proportional to the Bondi/Hill radius in the low-/high-mass regimes are consistent with the results for the isothermal and inviscid cases in \cite{f19}.
The transition planetary mass between the two regimes we found is roughly consistent with the expectation of \citet{f19} ($\hat{r}_{\rm H}\simeq 0.7$; $q_{\rm thermal}\simeq 4$ in their notation).

\subsection{Gas accretion onto circumplanetary disks}\label{sec:gas_accretion}
Accretion of small particles onto a CPD is largely influenced by the flow pattern of the gas around the planet. Previous studies show that the gas coming from a limited range of radial and vertical region in the PPD can accrete onto the CPD \citep[e.g.,][we call this region "accretion band", hereafter]{t12}. 
The initial radial location of dust particles accreting onto a CPD is related to gas accretion band \citep{h20}.
In this section, we focus on the radial and vertical structure of the gas accretion band, which is important for accretion of gas and dust particles into the CPD. 
Note that the structure of the gas accretion band is hardly influenced by the gap depth, which we cannot precisely evaluate by our local simulation.

\begin{figure}[h]
\begin{tabular}{c}

	\begin{minipage}{1\hsize}	
	\begin{center}
		\includegraphics*[bb=0 0 334 124,scale=0.8]{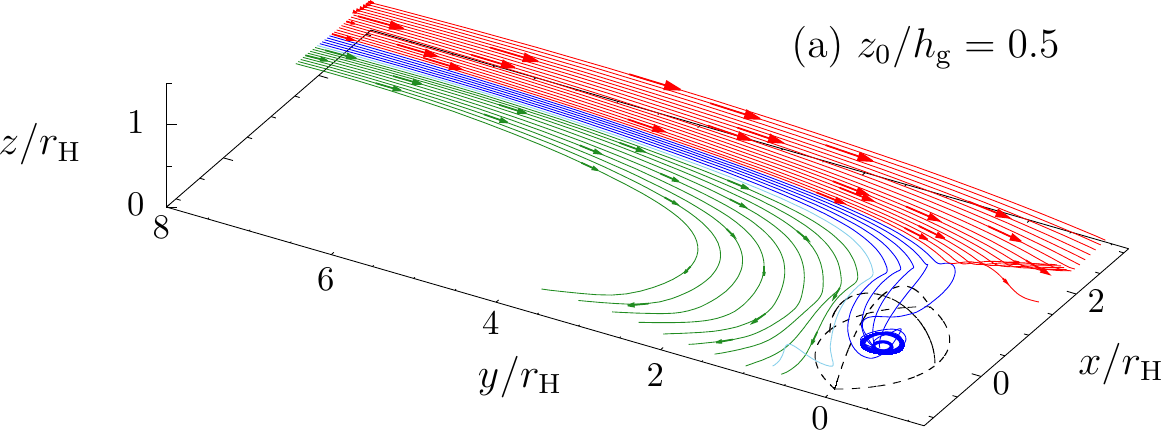}
	\end{center}
	\end{minipage}\\
\\	
	
	\begin{minipage}{1\hsize}	
	\begin{center}
		\includegraphics*[bb=0 0 334 116,scale=0.8]{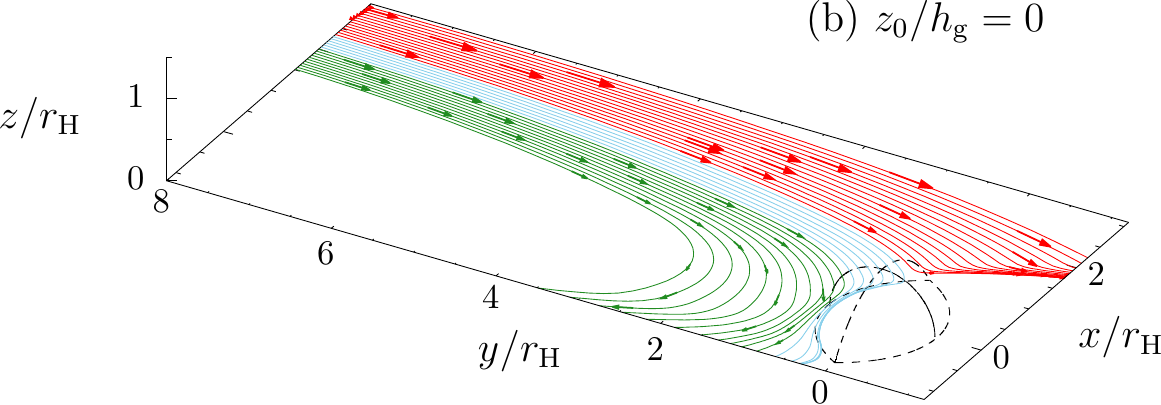}
	\end{center}
	\end{minipage}\\
	
\end{tabular}
	\caption{Classification of streamlines in the cases of (a) $z_0/h_{\mr{g}}=0.5$ and (b) $z_0/h_{\mr{g}}=0$ for the model with $\hat{r}_{\rm H}=1.36$ ($M_{\mr{p}}=1M_{\mr{Jup}}$). Colors show the types of streamlines. Red: passing (Keplerien shear flow), green: horseshoe (moving along a horseshoe orbit), dark-blue: accreting (reaching within $r=0.2r_{\mr{H}}$), light-blue: recycling (streamlines which enter the planet's Hill sphere and then escape from it). The lemon-shaped region enclosed by the dashed line in each panel is the planet's Hill sphere (its surface is defined by $\hat{\Phi}=0$ in \Equref{eq:Phi}).}
	\label{fig:stream}
\end{figure}
\begin{figure}[h]
	\begin{center}
		\includegraphics*[bb=0 0 256 229,scale=1]{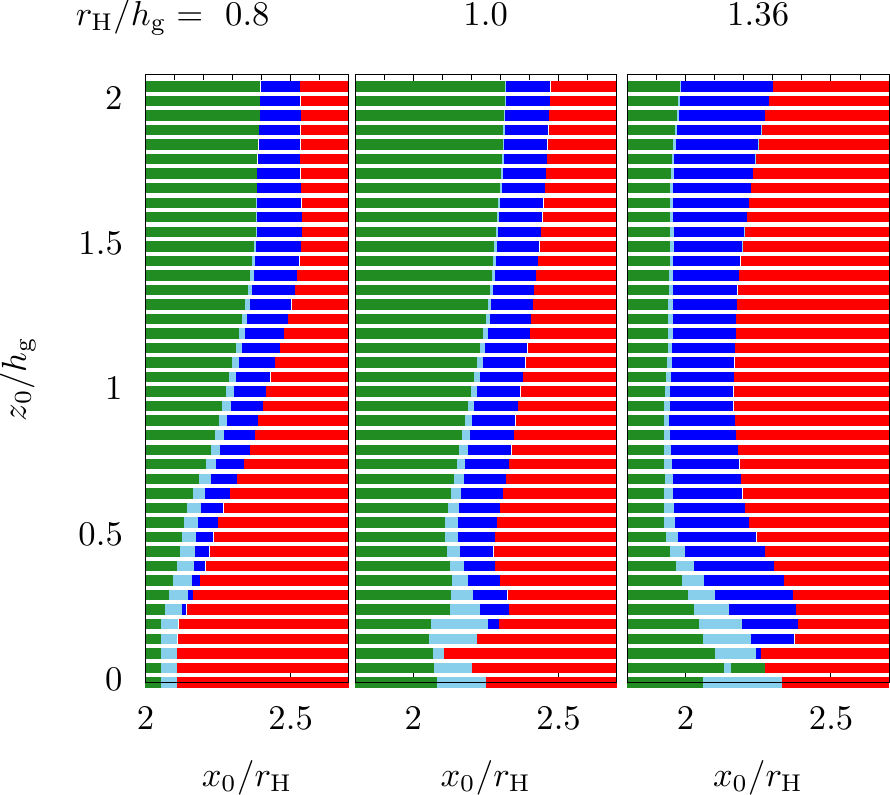}
	\end{center}
	\caption{Classification of initial positions ($x_0$, $z_0$) of streamlines of the gas flow at $y/h_{\rm g} = 12$ and $x > 0$ by their final destinations. The initial radial position $x_0$ is scaled by $r_{\rm H}$, while $z_0$ is scaled by the gas scale height $h_{\mr{g}}$. 
	Colors show the types of streamlines. Red: passing (Keplerien shear flow), green: horseshoe (moving along a horseshoe orbit), dark blue: accreting (reaching within $r=0.2r_{\mr{H}}$; we call this region accretion band), light blue: recycling (streamlines which enter the planet's Hill sphere and then escape from it).
	Three cases with $\hat{r}_{\rm H}=0.8$, 1.0, 1.36 ($M_{\mr{p}} = 0.2$, 0.4, and $1M_{\mr{Jup}}$) are shown.}
	\label{fig:bz_mp}
\end{figure}

In \Figrefs{fig:stream} and \ref{fig:bz_mp}, starting radial and vertical positions of streamlines ($x_0$ and $z_0$, respectively) at $\hat{y} = 12$ are classified by color according to their final destination \citep[][]{t12,h20}.
Those gas elements with large $x_0$ pass by the planet (passing; shown in red), while those with small $x_0$ move along horseshoe orbits (horseshoe; shown in green).
Some of the streamlines between passing and horseshoe regions reach within $r=0.2r_{\rm H}$ (accreting; shown in dark blue)\footnote{For the low-mass cases, we found that the gas streamlines which reach within $r=0.2r_{\rm H}$ also reaches within $r=0.07r_{\rm B}$ ($<0.2r_{\rm H}$) and the same accretion bands are obtained for both criteria of accretion. This means that the gas which reaches within $r=0.2r_{\rm H}$ loses its angular momentum at the shock surface on the CPD and flows toward the inner region of the CPD \citep[see also Figure 15 in][]{t12}.}, and others enter the planet's Hill sphere (whose surface is defined by the equipotential surface with $\hat{\Phi}=0$ in \Equref{eq:Phi}) and then escape from it (recycling; shown in light blue).
We confirmed the three-dimensional structure of gas accretion, which was reported by previous works that examined cases for $M_{\mr{p}}\sim M_{\mr{Jup}}$ \citep{m08,m10,t12,g13,sz14,s20} as well as lower-mass ($\sim$ Super-Earth) cases \citep{w14,k19,f19,l19}.
\Figref{fig:stream} shows the streamlines with (a) $\hat{z}_0=0.5$ and (b) $\hat{z}_0=0$, respectively, for $\hat{r}_{\rm H}=1.36$ ($M_{\mr{p}}=1M_{\mr{Jup}}$). 
The off-midplane gas elements encounter the planetary shock and change their direction slightly upward. Then, they accrete onto the CPD (\Figref{fig:stream}a). 
On the other hand, the gas elements which originate from the midplane cannot enter the Hill sphere, because radially outward flow dominates at the midplane \citep[\Figref{fig:stream}b; see also Figure 2 of][]{t12}.
It should be noted that the path of the accreting streamline likely depends on temperature structure. According to global radiative-hydrodynamic simulations \citep{s20}, the spiral arm tilt, which depends on the vertical structure of temperature, controls the direction of accreting gas flow. 
The direction change observed in \Figref{fig:stream}a is less clear than the one found in \citet{s20}. This is probably due to the difference in the temperature distribution. In our isothermal simulation, the vertical distribution of temperature does not have a gradient, which likely affects the tilt of the spiral arms.

\Figref{fig:bz_mp} shows radial and vertical distributions of each type of streamline at the starting points $(x_0, z_0)$ for planetary masses of $\hat{r}_{\rm H}=0.8$, 1.0, and 1.36 ($M_{\mr{p}}=0.2$, 0.4, and 1$M_{\mr{Jup}}$), respectively. In all these cases, we found that the accretion bands (the dark blue regions) do not exist near the midplane ($z_0\simeq 0$) but are distributed above the midplane continuously in the $z$-direction in a zonal pattern.
We also found that the radial ($x$-direction) width of the accretion band increases as the planetary mass increases. This indicates that a more massive planet can acquire gas and dust particles from a radially wider region in the PPD. 

Next, we evaluate the radial width of the accretion band and its dependence on planetary mass.
Let the non-dimensional radial width of accretion band at a scaled starting altitude $\hat{z}_0$ be $\hat{w}(\hat{z}_0)$. 
Assuming the hydrostatic equilibrium in the $z$-direction, the vertical density distribution of gas is given by
\begin{equation}
    \hat{\rho}(z)=\frac{1}{\sqrt{2\pi}} \exp \left( -\frac{\hat{z}^2}{2} \right).
\end{equation}
We take an average of $\hat{w}(\hat{z}_0)$ weighted by the above vertical density distribution as
\begin{equation}\label{eq:bar_w}
\bar{\hat{w}}=\frac{\int^{\hat{z}_{0,\mr{max}}}_0 \hat{w}(\hat{z}_0)\exp \left( -\frac{\hat{z}_0^2}{2} \right)d\hat{z}_0 }{\int^{\hat{z}_{0,\mr{max}}}_0 \exp \left( -\frac{\hat{z}_0^2}{2} \right)d\hat{z}_0},
\end{equation}
where we assumed $\hat{z}_{0,\mr{max}}=3$ because contribution of gas accretion from $\hat{z}_0>3$ is negligible.

\begin{figure}[H]
	\begin{center}
		\includegraphics*[bb=0 0 256 244,scale=1]{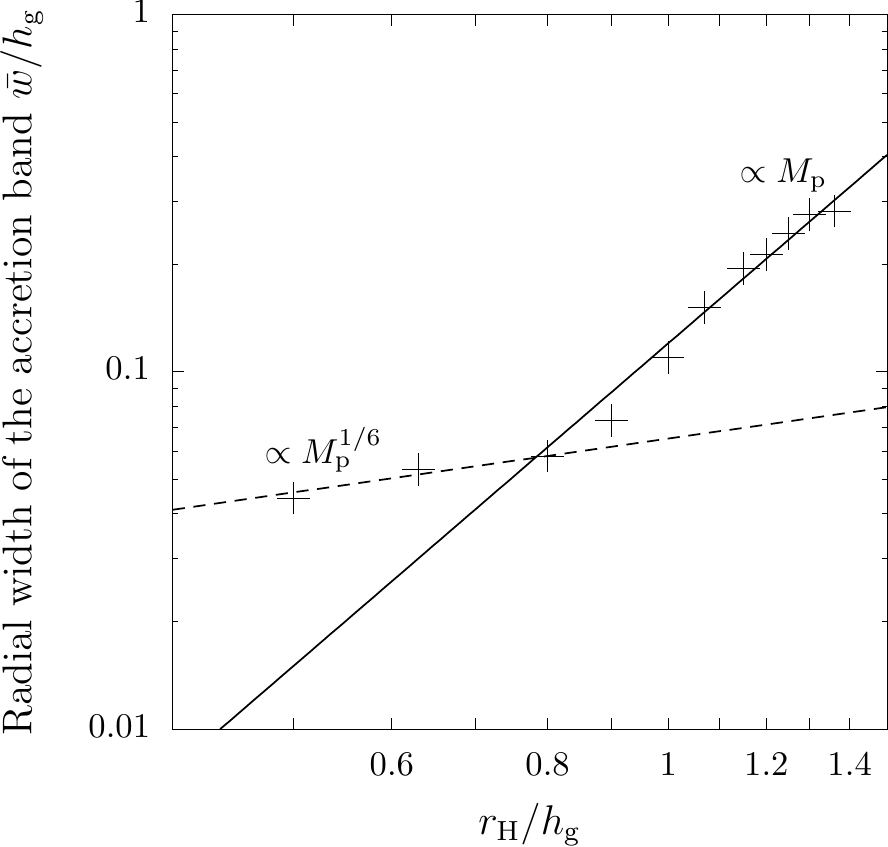}
	\end{center}
	\caption{Dependence of vertically averaged width of the accretion band $\bar{w}/h_{\mr{g}}$ on $\hat{r}_{\rm H}=r_{\mr{H}}/h_{\mr{g}}$. The dashed and solid lines show the functions denoted by \Equrefs{eq:w-mp_low} and (\ref{eq:w-mp_high}), respectively.}
	\label{fig:mp-width}
\end{figure}

\Figref{fig:mp-width} shows the dependence of the averaged width of the accretion band as a function of $\hat{r}_{\mr{H}}$, which depends on the planetary mass as shown in \Equref{eq:hat_rH}. 
The crosses are obtained from our numerical results. The dashed line is the fitting function for the low-mass regime:
\begin{equation}\label{eq:w-mp_low}
\bar{\hat{w}}=0.065 \hat{r}_{\mr{H}}^{1/2} = 0.054\left( \frac{M_{\mr{p}}}{M_{\mr{c}}} \right)^{1/6} \left( \frac{a}{h_{\mr{g}}} \right)^{1/2},
\end{equation}
and the solid line shows the fitting function for the high-mass regime:
\begin{equation}\label{eq:w-mp_high}
\bar{\hat{w}}=0.12 \hat{r}_{\mr{H}}^3 = 0.04\left( \frac{M_{\mr{p}}}{M_{\mr{c}}} \right) \left( \frac{a}{h_{\mr{g}}} \right)^3.
\end{equation}

Thus, planetary-mass dependences of the averaged accretion band width are different between the low-mass and high-mass regimes; $\bar{w} \propto r_{\mr{H}}^{1/2} \propto M_{\mr{p}}^{1/6}$ for $\hat{r}_{\mr{H}}\leq 0.8$ ($M_{\mr{p}}\leq 0.2 M_{\mr{Jup}}$) and $\bar{w} \propto r_{\mr{H}}^3 \propto M_{\mr{p}}$ for $\hat{r}_{\mr{H}}\geq 0.8$ ($M_{\mr{p}}\geq 0.2 M_{\mr{Jup}}$), respectively.
Note that the planetary-mass dependence of $\bar{w}$ for the low-mass regime has larger uncertainties due to the small number of cases we examined.

\subsection{Mass accretion rate onto circumplanetary disks}
\subsubsection{Vertical structure of accreting gas}
As shown in the previous section, the gas accreting into the CPD comes from high altitudes, and the vertical structure of the accretion band depends on the planetary mass. On the other hand, the gas density in the unperturbed PPD decreases with increasing altitude. In order to understand which vertical part of the PPD contributes most to the accretion into the CPD, we examine the mass accretion rate as a function of the initial altitude $\hat{z}_0$ of the accreting gas.

\begin{figure}[H]
	\begin{center}
		\includegraphics*[bb=0 0 300 245,scale=0.85]{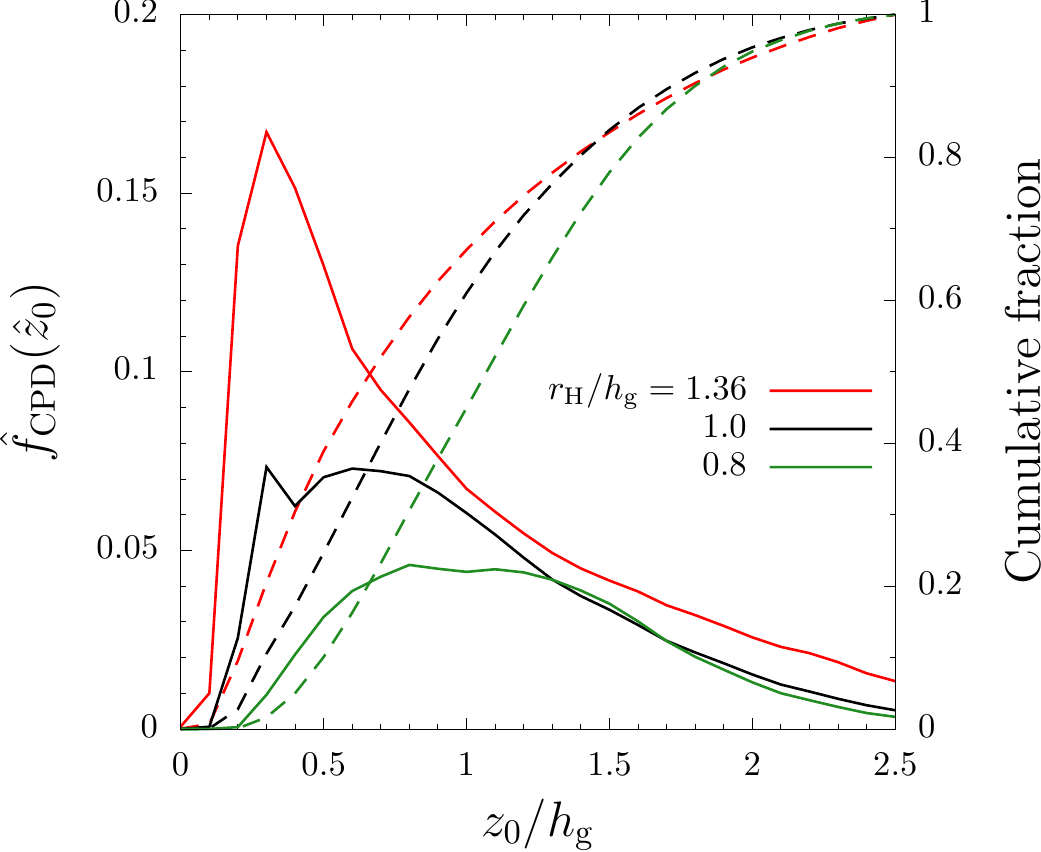}
	\end{center}

	\caption{Mass accretion rate per unit length in the $z$-direction of the gas originates from an altitude $\hat{z}_0$ into the CPD defined by \Equref{eq:f}. Only the gas accreting from $x_0>0,\; z_0>0$ is considered. The dashed lines show the cumulative distribution, which is normalized by the value at $z_0/h_{\mr{g}}=2.5$. Colors show the different planetary mass case; green shows $\hat{r}_{\mr{H}}=0.8$ (0.2$M_{\mr{Jup}}$), black shows $\hat{r}_{\mr{H}}=1.0$ (0.4$M_{\mr{Jup}}$), and red shows $\hat{r}_{\mr{H}}=1.36$ (1$M_{\mr{Jup}}$), respectively.}
	\label{fig:mass_z}
\end{figure}

\Figref{fig:mass_z} shows the mass accretion rate per unit length in the $z$-direction of the gas originating from an altitude $\hat{z}_0(>0)$ into the CPD defined as
\begin{equation}\label{eq:f}
\hat{f}_{\mr{CPD}}(\hat{z}_0)=\int^{\hat{x}_{0,\mr{max}}(\hat{z}_0)}_{\hat{x}_{0,\mr{min}}(\hat{z}_0)} \hat{\rho}(\hat{x}_0, \hat{z}_0) \hat{v}(\hat{x}_0, \hat{z}_0) d\hat{x}_0,
\end{equation}
where $\hat{x}_{0,\mr{min}}$ and $\hat{x}_{0,\mr{max}}$ are the values of $\hat{x}_0$ at the inner and outer boundaries of the accretion band at $\hat{z}_0$, respectively. \Equref{eq:f} includes only the gas accreting from $x_0>0$ and $z_0>0$.
We show the plots of $\hat{f}_{\mr{CPD}}(\hat{z}_0)$ for three cases of different planetary masses corresponding to $\hat{r}_{\mr{H}} = 0.8$, 1.0, and 1.36.
We found that a major contribution to the mass accretion comes from lower altitudes for higher planetary masses; the peak of the mass accretion rate is at $z_0/h_{\mr{g}}\simeq 1.0$, 0.6, and 0.3 for $\hat{r}_{\mr{H}}=0.8$, 1.0, and 1.36, respectively. This reflects the vertical structure of the accretion band (see \Figref{fig:bz_mp}), which becomes wider at lower $z_0$ and is distributed down to lower altitudes with increasing planetary mass. 
We also show in \Figref{fig:mass_z} the cumulative distribution of the mass accretion rate normalized by the total mass accretion rate with the dashed lines.
We found that about 80\% of gas accreting into the CPD comes from $z_0\lesssim 1.5h_{\mr{g}}$, 
and the contribution from lower altitudes increases with increasing planetary mass, as we described above. 
For example, the contribution from $z \leq 0.3 h_{\mr{g}}$ is $\simeq 1.6\%$, $\simeq 10\%$, and $\simeq 20\%$ for $\hat{r}_{\mr{H}} = 0.8$, 1.0, and 1.36, respectively. 

\subsubsection{Mass accretion rate evaluated from our simulation}
As shown in \Figref{fig:bz_mp}, only part of the gas entering the planet's Hill sphere comes sufficiently close to the planet to accrete into the CPD, and the rest is expelled back to the PPD as part of the meridional circulation \citep[e.g.,][]{t12,m14,fc16}.
Thus, the mass accretion rate of the gas into the CPD ($\dot{M}_{\rm acc, CPD}$) is generally smaller than that into the Hill sphere ($\dot{M}_{\rm acc, Hill}$), and the ratio $\dot{M}_{\rm acc, CPD}/\dot{M}_{\rm acc, Hill}$ is an important quantity when considering the delivery of small dust particles coupled to the gas into the CPD as building blocks of satellites.
Here we calculate $\dot{M}_{\mr{acc,CPD}}$ and $\dot{M}_{\mr{acc,Hill}}$ from the above analysis of the streamlines and using the following equations:
\begin{equation}\label{eq:Mcpd}
    \dot{\hat{M}}_{\rm acc, CPD} = 4 \int^{\hat{z}_{\mr{0,max}}}_0 \hat{f}_{\mr{CPD}}(\hat{z}_0) d\hat{z}_0
\end{equation}
and
\begin{equation}\label{eq:Mhill}
    \dot{\hat{M}}_{\rm acc, Hill} = 4 \int^{\hat{z}_{\mr{0,max}}}_0 \hat{f}_{\mr{Hill}}(\hat{z}_0) d\hat{z}_0,
\end{equation}
where $\hat{f}_{\mr{Hill}}(\hat{z}_0)$ is the mass accretion rate into the Hill sphere per unit length in the $z$-direction, which is obtained by using both recycling and accreting streamlines in a similar calculation to \Equref{eq:f} (see \Figrefs{fig:stream} and \ref{fig:bz_mp}). The factor of four is multiplied in \Equref{eq:Mcpd} and \Equref{eq:Mhill} to account for the symmetry about the $x=0$ and $z=0$ planes.
We did not adopt the sink cell. Thus, in the quasi-steady state of the gas flow in our simulations, the accreted gas accumulates in the CPD while part of it is expelled back toward the protoplanetary disk as an outflow near the midplane of the CPD (the top panel of \Figrefs{fig:vr_high} and \ref{fig:vr_low}). In order to precisely derive the amount of the gas that is accreted in the CPD but then expelled back toward the protoplanetary disk as an outflow from the midplane region of the CPD, we will need to investigate the long-term viscous evolution of the CPD. However, it is difficult to investigate it in our present simulation where the gas is assumed to be inviscid and is beyond the scope of the present work. Thus, in the present work, we derived the accretion rate onto the CPD from the analysis of the streamlines that reach within $r = 0.2 r_{\rm H}$ (i.e., the dark blue region in \Figref{fig:bz_mp}).

\begin{figure}[H]
	\begin{center}
		\includegraphics*[bb=0 0 211 261,scale=1.2]{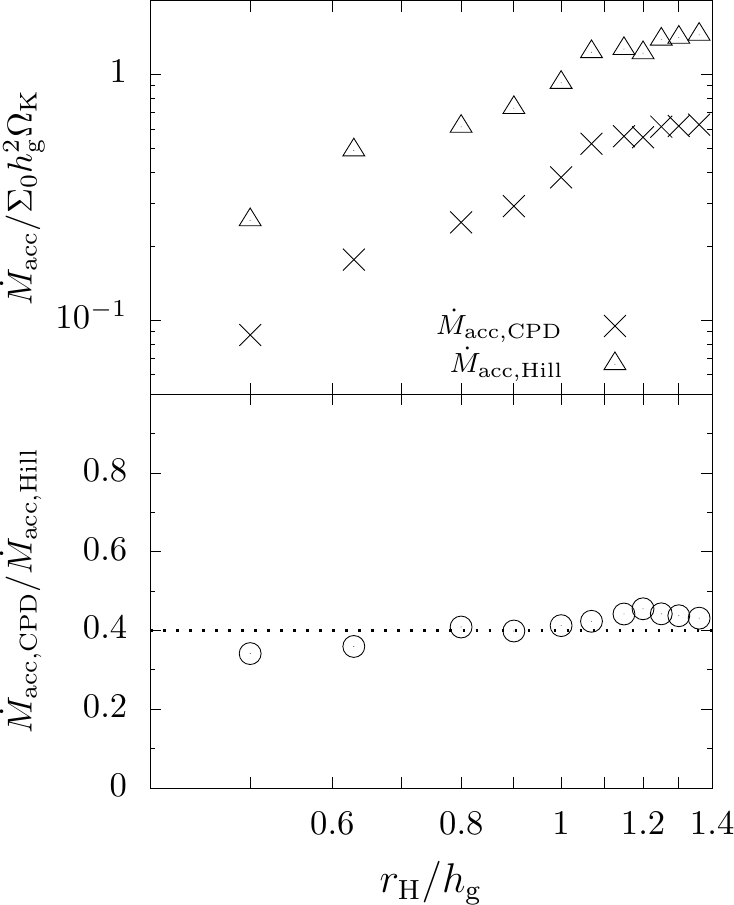}
	\end{center}

	\caption{Top panel: Dependence of non-dimensional mass accretion rates into the CPD ($\dot{M}_{\mr{acc,CPD}}$; crosses) and the Hill sphere ($\dot{M}_{\mr{acc,Hill}}$; triangles) on $\hat{r}_{\rm H}=r_{\mr{H}}/h_{\mr{g}}$. Bottom panel: The ratio $\dot{M}_{\mr{acc,CPD}}/ \dot{M}_{\mr{acc,Hill}}$ as a function of $\hat{r}_{\rm H}$. The dotted line represents $\dot{M}_{\mr{acc,CPD}}/ \dot{M}_{\mr{acc,Hill}}=0.4$.}
	\label{fig:Macc}
\end{figure}

In \Figref{fig:Macc}, we show the planetary mass dependence of $\dot{M}_{\rm acc, CPD}$ and $\dot{M}_{\rm acc, Hill}$ (top panel), and the ratio $\dot{M}_{\rm acc, CPD}/\dot{M}_{\rm acc, Hill}$ (bottom panel). 
We found that $\dot{M}_{\rm acc, Hill}/\Sigma_0 h_{\rm g}^2 \Omega_{\rm K} \simeq 0.4$ when $\hat{r}_{\rm H}=1$, which is consistent with Tanigawa et al. (2012, their Figure 14), who found that $\dot{M}_{\rm acc, Hill}/\Sigma_0 h_{\rm g}^2 \Omega_{\rm K} \simeq 0.2$ for the mass accretion rate onto the $z>0$ surface of the CPD for $\hat{r}_{\rm H}=1$.
We also found that $\dot{M}_{\mr{acc,CPD}}/ \dot{M}_{\mr{acc,Hill}} \simeq 0.4$ regardless of the planetary mass.
This means that gas accretion rate onto the CPD can be estimated from that into the Hill sphere in global simulation at least in the planetary mass range corresponding to $\hat{r}_{\mr{H}}=0.5-1.36$, as long as the resolution of the simulation is sufficient to resolve the gas flow on the Hill-radius scale.
Also, the fact that the ratio $\dot{M}_{\mr{acc,CPD}}/ \dot{M}_{\mr{acc,Hill}}$ is nearly independent of the planetary mass indicates that this ratio is insensitive to whether the gap opens or not, probably because it is determined by the distributions of the accretion band and recycling region (\Figref{fig:bz_mp}), which is likely to be insensitive to the gap depth.
It should be noted that absolute values of the mass accretion rates shown in \Figref{fig:Macc} depend on the gap depth at the planet's orbit, which depends on numerical settings such as simulation time and resolution \citep[e.g.,][]{k17}. 
Thus, in \Secref{sec:Macc_ana}, we will discuss the corrected values of mass accretion rate using the gap model obtained from global hydrodynamic simulations \citep{k15}. 

\newpage
\section{Discussions}\label{sec:discussions}
\subsection{Semi-analytical formulae of mass accretion rate}\label{sec:Macc_ana}
Here, we consider a physical interpretation of the planetary-mass dependence of mass accretion rates onto the CPD and derive semi-analytical formulae by combining our results with the gap model obtained from global hydrodynamic simulations \citep{k15}, because our local simulation cannot reproduce the gap depth accurately as we mentioned before.

Let us consider the gas flowing through the area of the accretion band at $(x,y)=(x_{\mr{ab}}, L_y/2)$, where $x_{\mr{ab}}$ is a typical value of the $x$-coordinate of the accretion band.
In the coordinate system rotating with the planet, the velocity of the gas passing the area of the accretion band is $(3/2) x_{\mr{ab}}\Omega_{\mr{K}}$ assuming the Keplerian shear flow at $y=L_y/2$.
Then, we can express the mass accretion rate onto the CPD as
\begin{equation}\label{eq:Macc_model}
\dot{M}_{\mr{acc,CPD}} = D\Sigma_{\mr{g,ab}},
\end{equation}
where $\Sigma_{\mr{g,ab}}$ is the surface density of the gas at $(x,y)=(x_{\mr{ab}}, L_y/2)$, and $D$ is the "accretion area" of the protoplanetary disk per unit time \citep{tt16}.
\citet{tt16} used the expression for $D$ for the gas accretion onto a growing planet based on the isothermal and inviscid two-dimensional hydrodynamic simulation of \citet{tw02}.
\footnote{\citet{tw02} defined the mass accretion rate by the time-averaged mass flux across the inner sink boundary at $r=0.05r_{\rm H}$.}
Here we assume that $D$ for the mass accretion rate on the CPD can be expressed in terms of the averaged accretion band width obtained from our three-dimensional hydrodynamic simulation (\Secref{sec:gas_accretion}) as
\begin{equation}\label{eq:D}
    D= 2\times \frac{3}{2}x_{\mr{ab}}\Omega_{\mr{K}} \bar{w}.
\end{equation}
The factor of two is multiplied in the r.h.s. of \Equref{eq:D} to account for the symmetry about $x=0$ plane.
For simplicity, we assume $x_{\mr{ab}}=2.2r_{\mr{H}}$, regardless of the planetary mass (\Figref{fig:bz_mp}). 

Substituting $x_{\rm ab}=2.2r_{\rm H}$ and the expression for $\bar{w}$ using \Equref{eq:w-mp_low} or (\ref{eq:w-mp_high}) into \Equref{eq:D}, we obtain 
\begin{equation}\label{eq:D_low}
    D = 0.25 \left( \frac{M_{\rm p}}{M_{\rm c}} \right)^{1/2} \left( \frac{a}{h_{\rm g}} \right)^{-1/2} a^2 \Omega_{\rm K} \qquad \mr{for}\; 0.5\leq \hat{r}_{\rm H} \leq 0.8 
\end{equation}
and
\begin{equation}\label{eq:D_high}
 D = 0.18 \left( \frac{M_{\rm p}}{M_{\rm c}} \right)^{4/3} \left( \frac{a}{h_{\rm g}} \right)^{2} a^2 \Omega_{\rm K}  \qquad \mr{for}\; 0.8\leq \hat{r}_{\rm H} \leq 1.36.
\end{equation}

\begin{figure}[H]
	\begin{center}
		\includegraphics*[bb=0 0 257 247,scale=0.8]{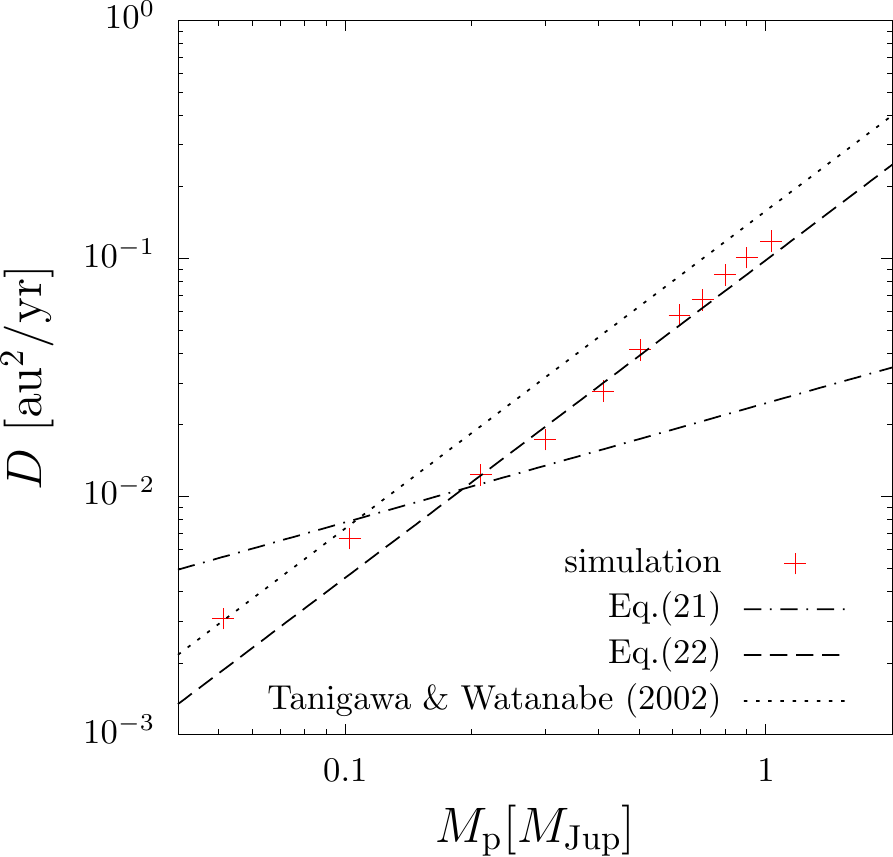}
	\end{center}

	\caption{Planetary-mass dependence of $D$ obtained by our hydrodynamic simulation ($=\dot{M}_{\rm acc,CPD}/\Sigma_{\rm g,ab,sim}$; red crosses) and the one calculated from \Equrefs{eq:D_low} (dot-dashed line) and (\ref{eq:D_high}) (dashed line). The values of $D$ obtained by two-dimensional simulation of \citet{tw02} are also plotted (dotted line). Following \citet{tt16}, we assume $a=5.2$~au.}
	\label{fig:D}
\end{figure}

In \Figref{fig:D}, we plot the values of $D$ calculated by \Equrefs{eq:D_low} and (\ref{eq:D_high}), and compare them with the results of our hydrodynamic simulation, where $D$ is derived by $\dot{M}_{\rm acc,CPD}/\Sigma_{\rm g,ab,sim}$ with $\Sigma_{\rm g,ab,sim}$ being the surface density at the accretion band obtained by our hydrodynamic simulation (\Figref{fig:Sg_ratio}). For comparison, we also plot the values of $D$ obtained from the two-dimentional simulation by \citet{tw02}, which was used in \citet{tt16}.
We found that the results of our hydrodynamic simulation can be well reproduced by \Equref{eq:D_high} in the high-mass regime, while they somewhat deviate from those calculated by \Equref{eq:D_low} in the low-mass regime.
This indicates that our model given by \Equrefs{eq:Macc_model} and (\ref{eq:D_high}) reproduces our numerical results well in the high-mass regime, but it is not satisfactory for the low-mass regime. The assumption that $x_{\rm ab}$ is independent of $M_{\rm p}$ may be too simple in the low-mass regime, and effects of different flow patterns for cases with $r_{\rm H}<h_{\rm g}$ would also be important.
In any case, it is difficult to construct an accurate model in the low-mass regime from our results, because in the present work we mainly focus on large-mass planets with a CPD in relation to satellite formation, and we ran only a few simulations for the low-mass regime. 
It should be noted that the planetary-mass dependence given by \Equref{eq:D_high} ($\propto M_{\rm p}^{4/3}$) is consistent with the one obtained by the two-dimensional simulation \citep{tw02}.
Although the qualitative behavior of accretion flow onto the CPD such as the strength of shocks is quite different between the two- and three-dimensional cases, the mass dependence of the accretion rates remains the same.

In the following, we assume that our simulation results can be approximated by \Equrefs{eq:Macc_model} and (\ref{eq:D_high}) for the whole range of planetary-mass we examined, and derive mass accretion rate using the empirical formula for the gap surface density obtained from global simulations \citep[][]{k15} given as 
\begin{equation}\label{eq:sigma_min}
 \frac{\Sigma_{\mr{min}}}{\Sigma_0} = \frac{1}{1+0.04K}
\end{equation}
with
\begin{equation}
\label{eq:K} K = \left( \frac{M_{\mr{p}}}{M_{\mr{c}}} \right)^2 \left( \frac{h_{\mr{g}}}{a} \right)^{-5} \alpha^{-1},
\end{equation}
where $\Sigma_{\mr{min}}$ is the surface density at the bottom of the gap.
We assume $\Sigma_{\mr{g,ab}}=\Sigma_{\mr{min}}$ because $x_{\mr{ab}}=2.2r_{\mr{H}}$ is located near the bottom of the gap.

\begin{figure}[H]
	\begin{center}
		\includegraphics*[bb=0 0 343 247,scale=0.8]{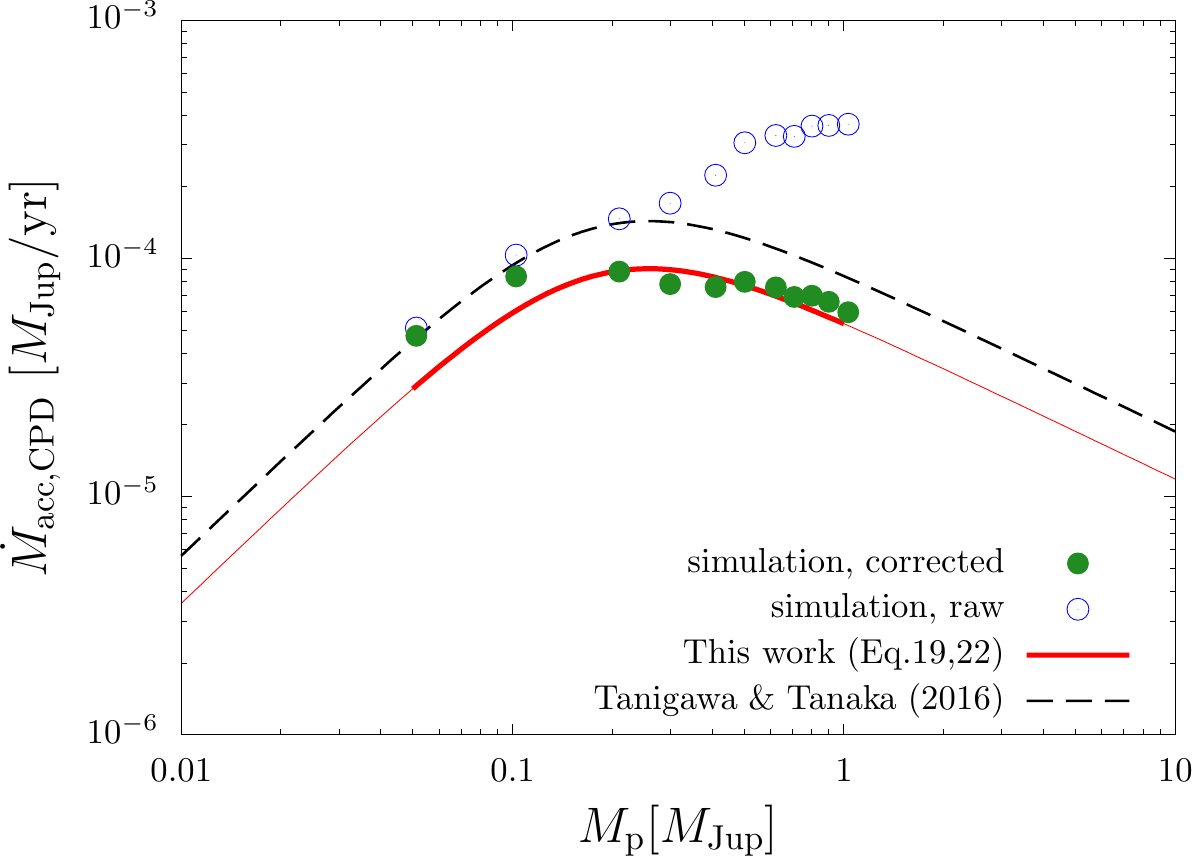}
	\end{center}

	\caption{Comparison of the mass accretion rates obtained by our hydrodynamic simulation (marks) and semi-analytical formulae (lines). The red line represents our semi-analytic formula (\Equrefs{eq:Macc_model} and (\ref{eq:D_high})), and the dashed line is that derived by \citet{tt16} using the results of two-dimensional hydrodynamic simulation. The raw values of our simulation results derived by \Equref{eq:Mcpd} are shown with the blue open circles, and those corrected by multiplying by $\Sigma_{\rm min}/\Sigma_{\rm g,ab,sim}$ are shown with the green filled circles. Thin lines are the extrapolation of our formulae beyond the mass range we examined with our simulation. $a=5.2$~au, $\Sigma_0=1.4\times 10^3 \; \mr{kg\; m^{-2}}$, and $\alpha=4\times 10^{-3}$ are assumed for direct comparison with Figure 1 in \citet{tt16}. }
	\label{fig:TT16}
\end{figure}

In \Figref{fig:TT16}, we plotted semi-analytical formula obtained from \Equref{eq:Macc_model} and  \Equref{eq:D_high} (red line), and our simulation results. 
Our simulation results derived by \Equref{eq:Mcpd} are shown with the blue open circles (raw values), and those corrected to reproduce the true gap depth by multiplying by $\Sigma_{\rm min}/\Sigma_{\rm g,ab,sim}$ (corrected values) are shown with the green filled circles.
We find that the semi-analytical formula for the mass accretion rate onto the CPD we obtained can reproduce the numerical results well in the high-mass regime, although they somewhat underestimate the numerical results in the low-mass regime.
In the high-mass regime, the planetary-mass dependence of our formula is consistent with the accretion rate onto the planet derived in \citet{tt16}, but the absolute values of the accretion rates derived from our three-dimensional simulation are about 60\% of those obtained in \citet{tt16}.

\citet{gc19} argued that, after the onset of runaway gas accretion but when the planet's Hill radius is still smaller than the disk scale height (i.e., $\hat{r}_{\rm H}<1$ in our simulations), the gas accretion rate is expected to match the Bondi accretion rate ($\propto M_{\rm p}^2$) rather than the empirical formula obtained from the results of hydrodynamic simulations \citep[$\propto M_{\rm p}^{4/3}$;][]{tt16} in the limit of low planetary mass ($r_{\rm B} \ll r_{\rm H}$), because the envelope boundary should be approximately spherically symmetric and the Keplerian shear velocity at $r_{\rm B}$ is negligible in such a case. 
It should be noted that our results derived on the basis of the analysis of the accretion bands obtained from hydrodynamic simulations cannot be applied to such low-mass cases.

\newpage

\subsection{Comparisons with previous works on gas accretion band}
Some previous works using high resolution local simulation also reported the increase of radial width of the gas accretion band as the planet mass increases.
\citet{tw02} performed two-dimensional hydrodynamic simulations and found that the radial width of the accretion band increases as the planet becomes more massive. Their detailed analysis of streamlines showed that energy dissipation at the spiral shock around the planet determines the radial width of the accretion band. The energy dissipation increases as the planet becomes more massive. 
However, our three-dimensional simulations did not show such significant energy dissipation at the spiral shock around the planet.
For the three-dimensional accretion, energy dissipation at the shock surface on the CPD is much stronger than that at the spiral arm of the planet \citep[see Figures 10 and 11 in ][]{t12}.
The structure of the accreting flow in the three-dimensional simulation is qualitatively different from the two-dimensional one. Further studies are needed to explain the planetary-mass dependence of accretion band width in the three-dimensional case with a simple model.

Although the planetary-mass dependence of accretion band width in the three-dimensional case is also examined by \citet{m10}, their definition of the accretion band width is different from ours. They defined the width by the difference between $x_{\mr{0,min}}$ and $x_{\mr{0,max}}$ defined for the whole accretion area in the $x_0-z_0$ plane, without measuring the radial width of the accretion band at each altitude.
The $z$-dependence of the accretion band width is important when discussing dust accretion onto the CPD \citep[e.g.,][]{h20} because vertical distribution of dust particles varies with turbulence strength in the PPD and deviates from gas distribution unless turbulence in the PPD is sufficiently strong. 

\subsection{Effects of thermodynamics}
In this study, we use the isothermal equation of state (EOS). In the late stage of planet formation, the gas surface density is expected to be low. 
Fluid in low-density regions mainly controls the accreting gas flow, and isothermal assumption should be valid in this region. We note that some regions near the midplane could be optically thick and not well-modeled by isothermal gas because the gap depth is moderately deep in our mass range (\Figref{fig:Sg_ratio} in Appendix).

\citet{f19} showed that the flow patterns are similar between isothermal and adiabatic cases when planetary mass is large enough ($\hat{r}_{\rm H}\geq 1.0$). This is because CPD dynamics are dictated by gravity, and the effect of EOS becomes small. For $\hat{r}_{\rm H}< 1.0$, on the other hand, the gravity of the planet is not so large, and EOS may affect the flow pattern around the planet. 
\citet{m10} investigated thermal effects on mass accretion rate and found that mass accretion rates with the isothermal and adiabatic EOSs are almost the same when planetary mass is small ($\hat{r}_{\rm H} \leq 1.0$). However, its planetary-mass dependence is slightly different for $\hat{r}_{\rm H}\geq 1.0$, and the isothermal case showed stronger dependence on planetary mass. This trend seems different from \citet{f19}.

As we mentioned in \Secref{sec:gas_accretion}, global radiation hydrodynamic simulations of \citet{s20} showed that the vertical temperature gradient causes the spiral arm tilt, which changes the direction of accreting flow. This indicates that the vertical temperature gradient likely affects the gas accretion onto CPD. Thus, analysis of the structure of the gas accretion band onto CPD with more realistic conditions is desirable \citep[e.g.,][]{sz16,sz17,f19}.


\subsection{Effects of Other Numerical Settings}
Settings of a numerical domain can affect the circumplanetary structure and gas accretion. Using global simulation, \citet{g13} reported that the CPD can have a non-axisymmetry structure, the side facing the central star being thicker. In contrast, the CPD of our local simulation is roughly axisymmetry around the planet. The non-axisymmetry in the global case may lead to different structures of the gas accretion bands in regions interior and exterior to the planet's orbit.
However, the basic characteristics of the structure of gas flow near the CPD (i.e., vertical accreting flow and outflow in the midplane) should be similar in both regions.

The radius of the CPD can depend on the effective viscosity (sum of the physical and the numerical viscosities). \citet{sz14} showed that the larger effective viscosity tends to produce a larger CPD. In our inviscid simulation, we found that the CPD radius is about $0.2r_{\mr{H}}$ for the massive planet cases and $0.08r_{\mr{B}}$ for the small planet cases, but these values may become larger for a viscous fluid. However, the qualitative results such as the CPD radius being proportional to the planet's Hill radius and the Bondi radius for massive and small planets, respectively, are likely not to change.

\section{Conclusions}\label{sec:conclusions}
We performed three-dimensional high-resolution hydrodynamic simulations to investigate the gas flow in the local region around the planet. The main purpose of the present work was to understand the planetary mass dependence of the accreting gas flow onto the CPD, because it is important for delivery of satellite building blocks to the CPD around planets with various masses and was not reported in the previous work \citep{t12}.
We performed simulations for $\hat{r}_{\rm H}=r_{\mr{H}}/h_{\mr{g}}=0.5-1.36$, corresponding to the planetary masses of $0.05M_{\mr{Jup}}-1M_{\mr{Jup}}$ at the current Jovian orbit for the MMSN model \citep{h81}.
The main results are as follows:
\begin{enumerate}
    \item The radius of the CPD is proportional to the planet's Hill sphere for the massive planet ($\hat{r}_{\mr{H}}\gtrsim 0.8$), while it is proportional to the Bondi radius for the small planet ($\hat{r}_{\mr{H}}\lesssim 0.8$). This is consistent with \citet{f19}.
    \item We examined the radial and vertical distribution of source points of streamlines that accrete onto the CPD. We found that source points of accreting streamlines are continuously distributed above the midplane in the vertical direction (we call this region accretion band). The radial width of the accretion band increases as the planetary mass increases. The width $\bar{w}$ is proportional to the planetary mass for the high-mass regime ($\hat{r}_{\mr{H}}\gtrsim 0.8$), while it is proportional to 1/6 power of the planetary mass for the small planet cases ($\hat{r}_{\mr{H}}\lesssim 0.8$).
    \item We found that the ratio of the mass accretion rate onto the CPD and that into the Hill sphere is about 0.4 regardless of the planetary mass. This means that the gas accretion rate onto the CPD can be estimated from that into the Hill sphere in global simulation as long as the resolution of the simulation is sufficient to resolve the gas flow on the Hill-radius scale.
    \item We derived semi-analytical formulae by combining our results with the gap model obtained from global hydrodynamic simulations \citep{k15}. Comparing with the semi-analytical formula obtained using two-dimensional simulation results \citep{tt16}, in the high-mass regime (${M}_{\mr{p}}\gtrsim 0.2M_{\rm Jup}$), we found that the values of the accretion rates derived from our three-dimensional simulation are about 60\% of those obtained from the two-dimentional case. In contrast, the mass dependence of the accretion rates remains the same between the two- and three-dimensional cases, although the qualitative behavior of accretion flow onto the CPD such as the strength of shocks is quite different between the two cases. We also found that the mass dependence of the accretion rate is different between the two- and three-dimensional cases for the low-mass regime (${M}_{\mr{p}}\lesssim 0.2M_{\rm Jup}$), but our semi-analytical model is too simple to reproduce the numerical results for the low-mass regime and further studies are needed to construct a more accurate model. 
\end{enumerate}

When dust particles are small and strongly coupled with the accreting gas, the mass accretion rate of dust particles should be strongly influenced by the gas.
Since actual motion of dust particles is expected to decouple from the gas \citep[e.g.,][]{h20}, orbital integration of particles is needed to study supply of dust onto CPD. We will report results of orbital calculation of dust particles in such cases in our subsequent paper.

In this work, we focused on planets whose mass is below the Jovian mass. However, more massive planets have been observed beyond the Solar system \citep[e.g.,][]{lj20}. Understanding satellite formation around exoplanet systems is important to construct a more general theory of satellite formation. Recently, some CPD candidates have been found \citep{g18,i19,t19,be21} as well as exomoon candidates \citep[e.g.,][]{tk18,k22}. Further studies of gas and dust accretion onto the CPD of planets with various masses will help us understand satellite formation in our Solar System as well as in exoplanet systems.

\begin{acknowledgments}
We thank Hidekazu Tanaka for helpful discussion. We also thank the anonymous reviewer for the careful report that helped us improve the manuscript.
T.T. sincerely appreciate Willy Kley for all of his energetic research activities, which had been motivating T.T. for decades.
This work was supported by
JSPS KAKENHI Grant numbers JP22J10202, JP18K11334, JP21H00043, JP22H01286, JP19K14787, JP15H02065, and JP20K04051. 
N.M. gives thanks for the support by the Kobe University
Doctoral Student Fellowship and JSPS Research Fellowship for Young Scientists.
This research used the computational resources of 
the 2020 and 2021 Koubo Kadai on Earth Simulator (NEC SX-ACE) at JAMSTEC, 
supercomputing resources at Cyberscience Center, Tohoku University, and 
the general-purpose PC cluster at the
Center for Computational Astrophysics, National Astronomical
Observatory of Japan.

\end{acknowledgments}


\appendix

\section{Gas Surface Density at the accretion band}\label{appendix}
\begin{figure}[H]
	\begin{center}
		\includegraphics*[bb=0 0 251 242,scale=0.9]{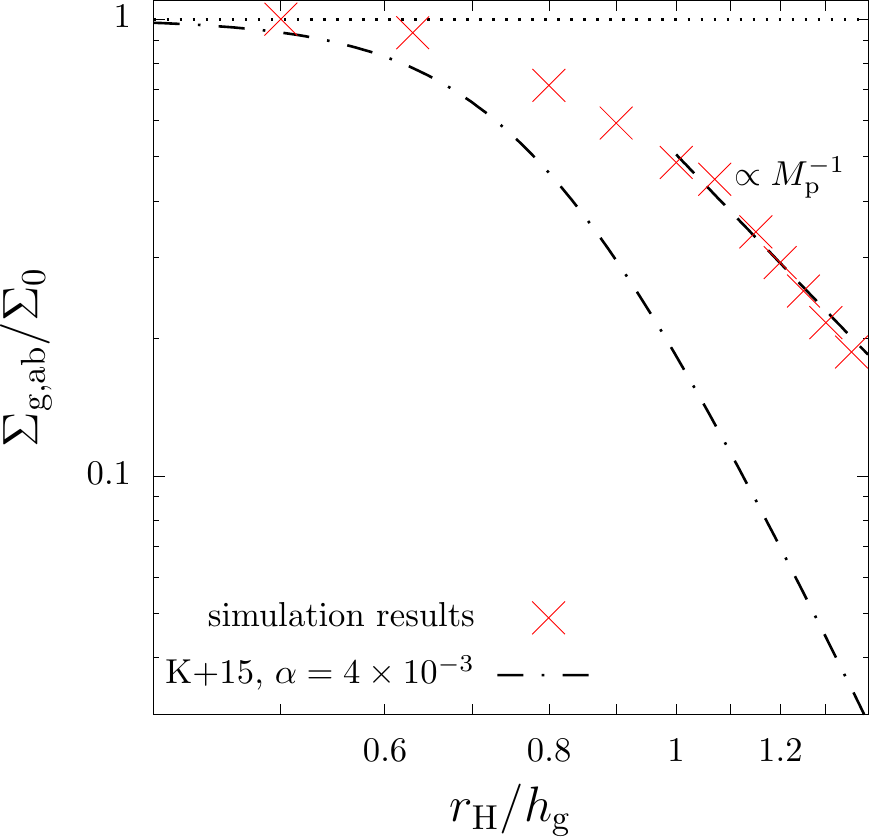}
	\end{center}

	\caption{Dependence of surface density at the accretion band $\Sigma_{\mr{g,ab}}$, i.e., surface density of gas at $(x,y)=(2.2r_{\mr{H}},L_y/2)$, on $\hat{r}_{\rm H}=r_{\mr{H}}/h_{\mr{g}}$. The dashed line shows the function proportional to $M_{\mr{p}}^{-1}$. The dotted-dashed curve shows surface density at the gap bottom derived by \citet{k15} for the case of $\alpha=4\times 10^{-3}$.}
	\label{fig:Sg_ratio}
\end{figure}

The planetary-mass dependence of the gas surface density at the accretion band, $\Sigma_{\mr{g,ab,sim}}$, is shown in \Figref{fig:Sg_ratio} with red crosses. $\Sigma_{\mr{g,ab,sim}}/\Sigma_0$ has values near unity in the low-mass limit, i.e., gap opening is little progressed in this regime. 
On the other hand, it has a planetary-mass dependence of $\Sigma_{\mr{g,ab,sim}}\propto M_{\mr{p}}^{-1}$ in the high-mass regime ($\hat{r}_{\mr{H}}\gtrsim 1.0$).
The deviation between our results and those obtained from the empirical formulae obtained by global simulation, \Equrefs{eq:sigma_min} and (\ref{eq:K}), (dot-dashed curve) increases with increasing planetary mass. This deviation is caused by the use of local simulation in the present work.

\end{document}